\documentclass[cits]{PoS}

\title{Signal-background interference in $gg \to H \to VV$}

\ShortTitle{Signal-background interference in $gg \to H \to VV$}

\author{\speaker{Nikolas Kauer}\\
Department of Physics, Royal Holloway, University of London, Egham TW20 0EX, UK\\
E-mail: \email{n.kauer@rhul.ac.uk}}

\abstract{The resonance-continuum interference between the SM Higgs search signal 
process $gg \to H \to VV$  ($V=W,Z$) and the irreducible background 
process $gg \to VV$ is studied at leading order for integrated cross sections 
and differential distributions in $pp$ collisions at $\sqrt{s}=7$ TeV and $14$ TeV
for $M_H=400$ GeV.  Leptonic weak boson decays are included, and realistic experimental selection cuts are applied.}

\FullConference{10th International Symposium on Radiative Corrections (Applications of Quantum Field Theory to Phenomenology)\\
		 September 26-30, 2011\\
		 Mamallapuram, India}

\usepackage{amsmath}

\newcommand{\sla}[1]{\ifmmode%
  \setbox0=\hbox{$#1$}%
  \setbox1=\hbox to\wd0{\hss$/$\hss}\else%
  \setbox0=\hbox{#1}%
  \setbox1=\hbox to\wd0{\hss/\hss}\fi%
  #1\hskip-\wd0\box1 }

\begin{document}

% =========================================================================

\section{Introduction}

\begin{figure}
\begin{center}
\includegraphics[height=2.1cm, clip=true]{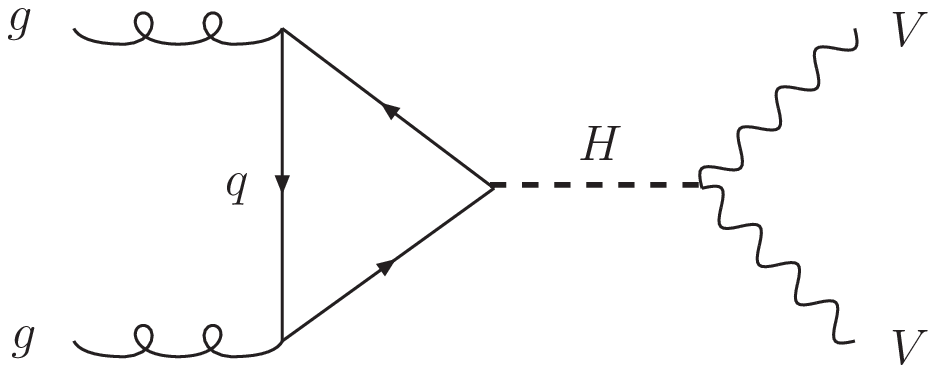}
\includegraphics[height=2.1cm, clip=true]{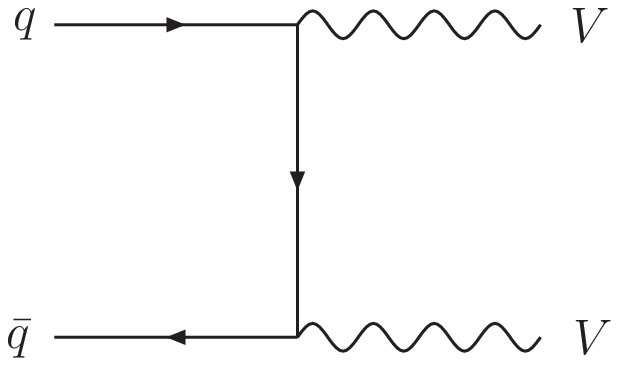}
\includegraphics[height=2.1cm, clip=true]{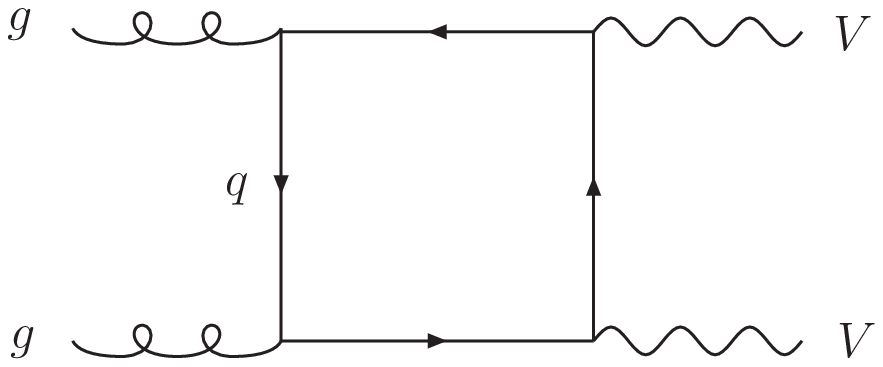}
\end{center}
\caption{Representative Feynman graphs for the Higgs signal process (left) and the $q\bar{q}$- (centre) and $gg$-initiated (right) continuum background processes at LO.}
\label{graphs}
\end{figure}

Higgs production in gluon fusion with subsequent decay into a 
weak boson pair is an important element of the Higgs search at the LHC.
The signal process (Fig.~\ref{graphs}, left) has been calculated and studied up to NNLO 
(see Refs.\ \cite{Djouadi:1991tka,Dawson:1990zj,Spira:1995rr,Harlander:2002wh,Anastasiou:2002yz,Ravindran:2003um,Anastasiou:2004xq,Anastasiou:2007mz,Catani:2007vq,Grazzini:2008tf,Anastasiou:2011pi} and references therein).
Continuum weak boson pair production is the dominant irreducible background
and has also been studied extensively.  For the leading quark scattering
subprocess (Fig.~\ref{graphs}, centre), programs at NLO are available, in part based on earlier calculations
(see Refs.\ \cite{Campbell:1999ah,Campbell:2011bn,Melia:2011tj} and references therein).
Here, we focus on the gluon scattering subprocess (Fig.~\ref{graphs}, right) and its interference 
with the signal process.  $gg\to VV$ continuum production formally enters
at NNLO and was calculated in Refs.\ \cite{Dicus:1987dj,Glover:1988fe,Glover:1988rg,Kao:1990tt}.  Off-shell weak boson decays and the possibility to interface with shower programs were subsequently included \cite{Campbell:2011bn,Zecher:1994kb,Binoth:2005ua,Duhrssen:2005bz,Binoth:2006mf,Binoth:2008pr,Campbell:2011cu,Frederix:2011ss}.  At the LHC, the importance of gluon-induced $VV$ continuum production and decay is enhanced by the large gluon-gluon flux and experimental Higgs search selection cuts.  
Resonance-continuum interference has been studied for $gg\ (\to H) \to VV$
in Refs.\ \cite{Anastasiou:2011pi,Glover:1988fe,Glover:1988rg,Binoth:2006mf,Campbell:2011cu,Seymour:1995qg} and for related processes in Refs.\ \cite{Dixon:2003yb,Dixon:2008xc,Boer:2011kf}.  Here, results for a heavy Higgs boson with $M_H=400$ GeV 
are presented.  The search for a heavy Higgs boson at hadron colliders
has been examined in Refs.\ \cite{Baur:1990af,Bagger:1993zf,Iordanidis:1997vs,Goria:2011wa}.

% =========================================================================

\section{Calculational details}

\begin{figure}
\begin{center}
\includegraphics[height=2.1cm, clip=true]{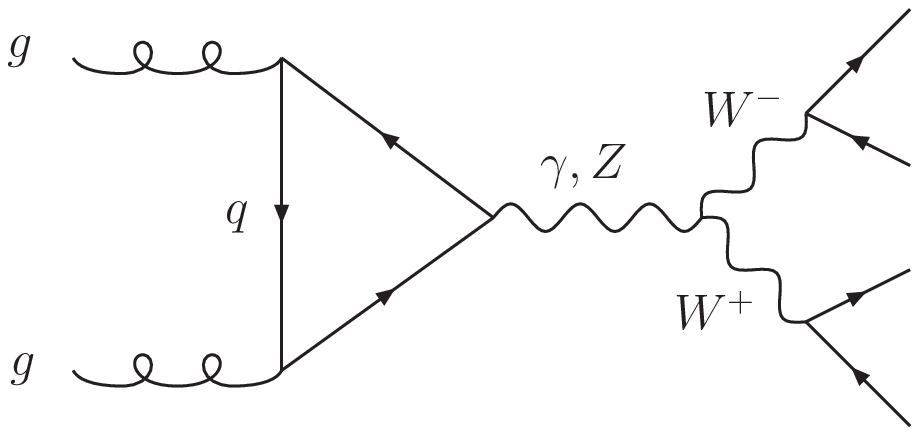}\hfil
\includegraphics[height=2.1cm, clip=true]{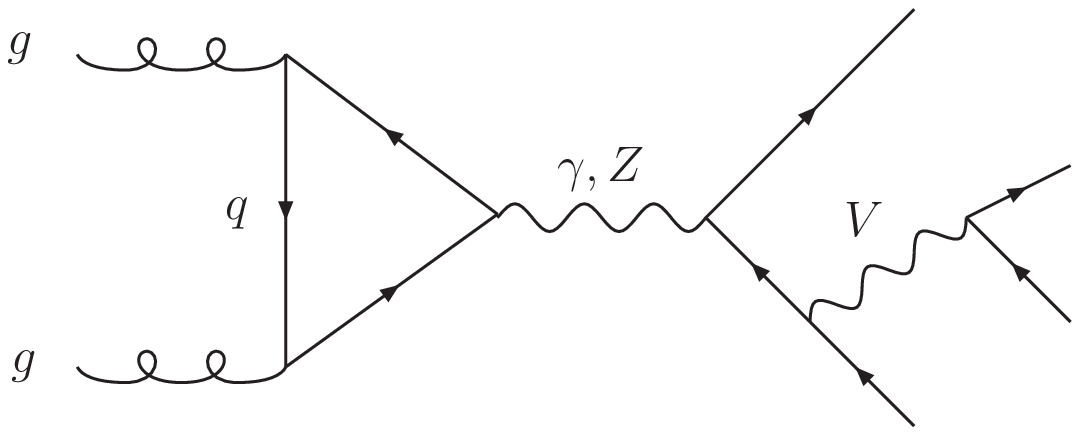}
\end{center}
\caption{Representative triangle graphs that formally contribute to gluon-induced $VV$ continuum production and decay.}
\label{triangles}
\end{figure}

The results presented in Section \ref{sec:results} have been calculated
with the programs gg2WW \cite{Binoth:2005ua,Binoth:2006mf} and gg2ZZ \cite{Binoth:2008pr}.  
Representative graphs for the $gg \to H \to VV$ signal process and the $gg \to VV$ continuum process are displayed in Fig.\ \ref{graphs}.  In addition
to box topologies in principle also triangle topologies 
contribute to the $gg$-initiated continuum processes (see Fig.\ \ref{triangles}).
But, in the limit of vanishing lepton masses the triangle graphs do not contribute.\footnote{Note that the $gg\to Z$ triangle graphs do contribute for non-zero lepton masses, which was verified by explicit calculation.}
For cross checks, two amplitude codes have been 
used based on the methods described in Ref.\ \cite{Binoth:2006mf} (BCKK) and 
Refs.\ \cite{Hahn:1998yk,Hahn:2000kx} (FormCalc).  Off-shell weak boson contributions and massless as well as massive quarks are taken into account.  Third quark generation contributions increase the total $gg\to WW\to$ leptons continuum cross section by $12\%$ and the double-resonant $gg\to ZZ\to$ leptons continuum cross section by $65\%$ ($pp$, $\sqrt{s}=14$ TeV).  For $Z$-pair production and decay, the $\gamma^\ast$ contributions are taken into account.  For $W$-pair production and decay, the BCKK
code approximates $V_\text{CKM} = 1$.  The FormCalc code was used to confirm
at the amplitude level that this is an excellent approximation as CKM effects are 
smaller than $0.01\%$.  
The gg2VV programs allow for the simultaneous calculation of
cross sections for multiple scales as well as the PDF error.\footnote{%
Sample output: scale1: 10.5817 MC: $\pm$0.0063 ($\pm$0.059\%) scale($\times$2): $-$2.5573 ($-$24\%) +3.6967 (+35\%) PDF: $-$0.2723 ($-$2.6\%) + 0.2382 (+2.3\%) fb, sym.\ scale error: $\pm$28\%, sym.\ PDF error: $\pm$2.4\%, scale2: 19.121 MC: $\pm$0.012 ($\pm$0.061\%) fb}

As both amplitude implementations employ Passarino-Veltman-type tensor 
integral reduction methods, a discussion of numerical stability is warranted.
The box amplitudes are affected by spurious singularities, which are caused by
inverse powers of the Gram determinant $\det G = 2s(tu - M^2_{V_1^\ast}M^2_{V_2^\ast})$, because $\det G\to 0$ as $p_{TV}\to 0$.  For phenomenologically relevant cross sections, the error caused by numerical instabilities should be small compared to practical MC integration errors, which are typically of order $0.1\%$.  After the symbolic cancellation of Gram determinants, the BCKK amplitude code evaluated in quadruple precision is numerically stable in the above sense.\footnote{Differential distributions are smooth, even when calculating extreme cross sections like $\sigma(p_{TV}<1\text{ GeV})$.}  On the other hand, numerical instability is observed for the FormCalc 
amplitude code when evaluated in quadruple precision for a relevant phase space (PS)
configuration with $p_{TV}=0.007$ GeV.  Using the stable BCKK code as benchmark, the 
following diagnostic algorithm was devised to detect problematic PS points with 
the FormCalc code and quadruple precision: First, we exploit that 
instabilities spoil Lorentz invariance by comparing $|{\cal M}|^2$ evaluated at 
the PS point and the PS point boosted along the beam axis with $p_{\text{boost}}=(1,0,0,0.001+0.1\,r_1)$ GeV with random $r_1\in[0,1]$.\footnote{The PS point is assumed to be given in the rest frame of $p_{\text{boost}}$. The PS point is boosted to the frame in which $p_{\text{boost}}$ is given.}  
The relative deviation is assessed 
using ${\rm reldev}(x,y) = |x-y|/\min(x,y)$. If ${\rm reldev}(|{\cal M}|^2,|{\cal M}_{\text{boosted}}|^2) > 10^{-4}$ the PS point is classified as unstable.
The same criterion is then applied again, except now with a random boost in 
the opposite direction.  If the PS point is still considered stable another
test is performed, which exploits that instabilities occur at exceptional PS configurations.  One therefore compares $|{\cal M}|^2$ evaluated at the PS point (in double precision) and the PS point mapped to single precision. If ${\rm reldev}(|{\cal M}|^2,|{\cal M}_{\text{single}}|^2) > 1$ the PS point is classified as unstable.
Unstable PS points are discarded.  Only in combination allow these tests to detect all instabilities.  As indicated above, this has been verified by comparison with the stable BCKK code.
One can thus also assess the error introduced by discarding PS points that
are wrongly classified as unstable.  The parameters have been adjusted 
to minimize this error.  For integrated cross sections, it is approximately 0.03\%.

% =========================================================================

\section{Results\label{sec:results}}

Parton-level cross sections for $gg\ (\to H)\to W^-W^+\to l\bar{\nu}_l\bar{l'}\nu_{l'}$ and $gg\ (\to H)\to ZZ \to l\bar{l}l'\bar{l'}$ ($l$: charged lepton) in $pp$ collisions at $\sqrt{s} = 7$ TeV are presented in Table \ref{tab:results}.  Results are given for a single lepton flavour 
combination, e.g.\ \,$l=e^-,l'=\mu^-$.  Lepton masses are neglected.
The input parameter set of Ref.\ \cite{Dittmaier:2011ti}, App.\ A, is used
with NLO $\Gamma_V$ and $G_\mu$ scheme.  For the Higgs resonance, we set
$M_H=400$ GeV and $\Gamma_H=29.16$ GeV \cite{Djouadi:1997yw}.
The renormalisation and factorisation scales are set 
to $M_H/2$.  The PDF set MSTW2008LO with 1-loop running for 
$\alpha_s(\mu^2)$ and $\alpha_s(M_Z^2)=0.13939$ is used. The fixed-width 
prescription is used for Higgs and weak boson propagators. For $ZZ$ production and decay, 
the virtual photon contributions have been included.

The following experimental selection cuts 
are adopted \cite{Dittmaier:2011ti,Chatrchyan:2011tz}:
As $WW$ standard cuts, we use 
$p_{Tl} > 20$ GeV, $|\eta_l| < 2.5$,
$\sla{p}_T > 30$ GeV and $M_{l\bar{l'}} > 12$ GeV.
As $WW$ Higgs search cuts for $M_H=400$ GeV, we use 
the $WW$ standard cuts and in addition 
$p_{Tl,\text{min}} > 25$ GeV, $p_{Tl,\text{max}} > 90$ GeV,
$M_{l\bar{l'}} < 300$ GeV and $\Delta\phi_{l\bar{l'}} < 175^\circ$.
As $ZZ$ standard cuts, we use
$p_{Tl} > 20$ GeV, $|\eta_l| < 2.5$ and
$76$ GeV $< M_{l\bar{l}}, M_{l'\bar{l'}} < 106$ GeV.
As $ZZ$ Higgs search cuts, we use the $ZZ$ standard cuts 
and in addition $|M_{l\bar{l}l'\bar{l'}} - M_H| < \Gamma_H$.

\begin{table}
\begin{center}
\renewcommand{\arraystretch}{1.2}
\begin{tabular}{|l|c|ccc|cc|}
\cline{3-7} 
\multicolumn{2}{c|}{} & \multicolumn{3}{|c|}{$\sigma$ [fb], $pp$, $\sqrt{s} = 7$ TeV, $M_H=400$ GeV} & \multicolumn{2}{c|}{interference} \\
\hline
\multicolumn{1}{|c|}{process} & cuts & $|{\cal M}_H|^2$  & $|{\cal M}_\text{cont}|^2$ & $|{\cal M}_H+{\cal M}_\text{cont}|^2$ & $R_1$ & $R_2$ \\
\hline
$gg\ (\to H) \to WW$ & stand. & 4.361(3) & 6.351(4) & 10.582(7) & 0.9879(8) & 0.970(2) \\
$gg\ (\to H) \to WW$ & Higgs  & 2.502(2) & 0.633(1) & 3.007(3) & 0.959(2) & 0.949(2) \\
$gg\ (\to H) \to ZZ$ & stand. & 0.3654(4) & 0.3450(4) & 0.7012(8) & 0.987(2) & 0.975(3) \\
$gg\ (\to H) \to ZZ$ & Higgs  & 0.2729(3) & 0.01085(2) & 0.2867(3) & 1.010(2) & 1.011(2) \\
\hline
\end{tabular}
\end{center}
\caption{Cross sections in fb for $gg\ (\to H)\to W^-W^+\to l\bar{\nu}_l\bar{l'}\nu_{l'}$ and $gg\ (\to H)\to ZZ \to l\bar{l}l'\bar{l'}$ in $pp$ collisions at $\sqrt{s} = 7$ TeV for $M_H=400$ GeV and a single lepton flavour combination calculated at LO.  Standard cuts and Higgs search cuts are applied (see main text).
Interference effects are illustrated through 
$R_1=\sigma(|{\cal M}_\text{VV}|^2)/\sigma(|{\cal M}_H|^2 + |{\cal M}_\text{cont}|^2)$ 
and $R_2=\sigma(|{\cal M}_H|^2+2\,\mbox{Re}({\cal M}_H{\cal M}^\ast_\text{cont}))/\sigma(|{\cal M}_H|^2)$.
}
\label{tab:results}
\end{table}

\begin{figure}
\begin{center}
\includegraphics[height=5.cm, clip=true]{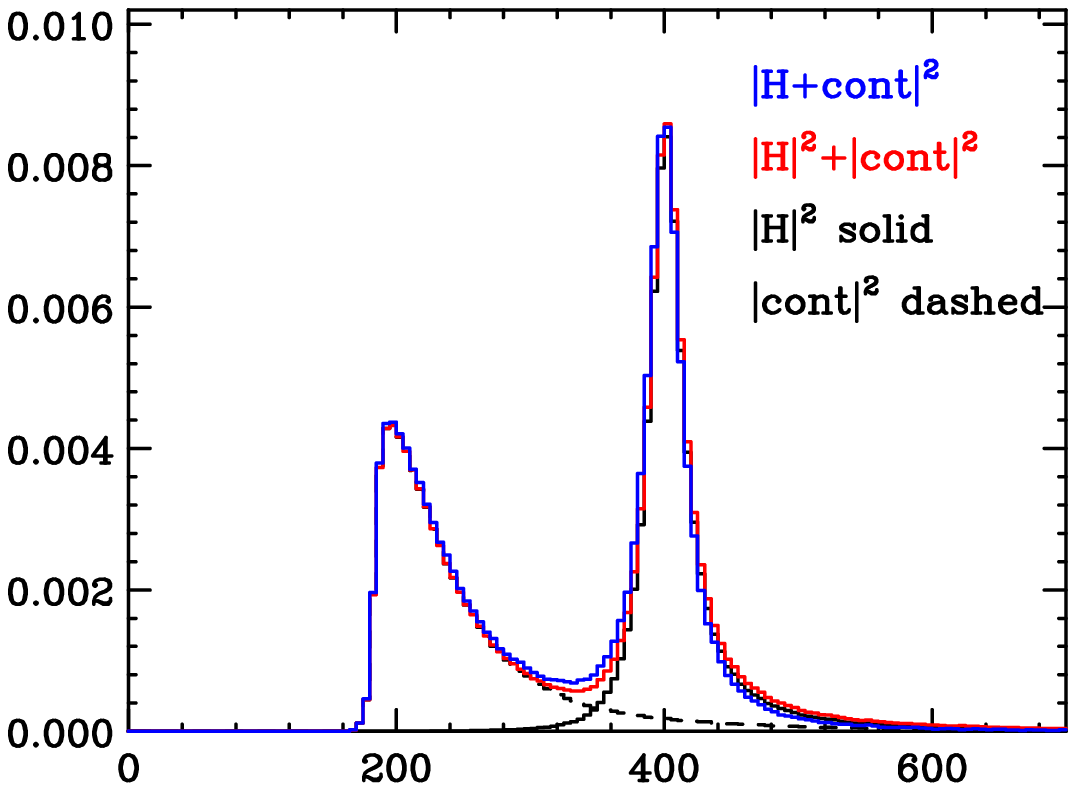}\hfil
\includegraphics[height=5.cm, clip=true]{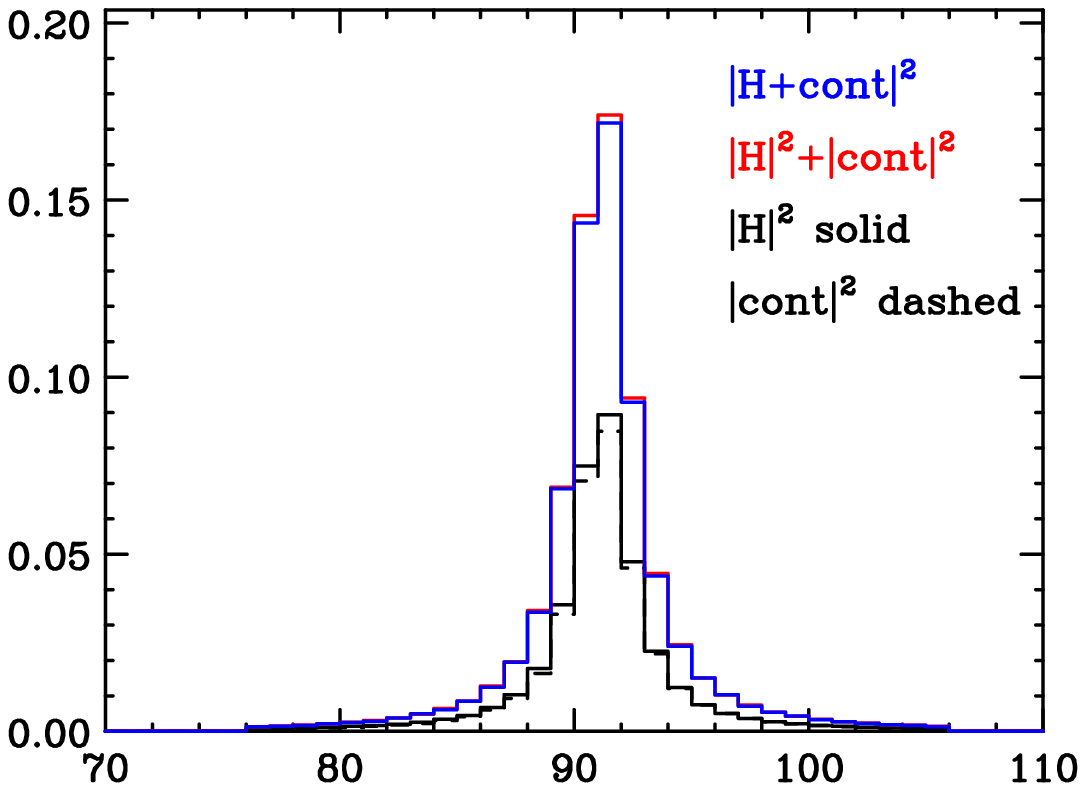}
\end{center}
\caption{Differential cross section distributions for $gg\ (\to H)\to ZZ \to l\bar{l}l'\bar{l'}$ in $pp$ collisions at $\sqrt{s} = 7$ TeV for $M_H=400$ GeV and a single lepton flavour combination calculated at LO.  The $M_{l\bar{l}l'\bar{l'}}$ [GeV] (left) and $M_{l\bar{l}}$ [GeV] (right) distributions are shown. Standard cuts are applied (see main text). fb is used as cross section unit.}
\label{fig:ZZ}
\end{figure}

The significance of a Higgs or new physics observation is determined
as function of the number of signal events 
$S\propto\sigma(|{\cal M}_\text{sig}|^2)$ and background events 
$B\propto\sum\sigma(|{\cal M}_\text{bkg}|^2)$.  When signal and 
background interfere the distinction becomes blurred.
Here: $|{\cal M}_\text{VV}|^2 = |{\cal M}_H|^2 + |{\cal M}_\text{cont}|^2 + 2\,\mbox{Re}({\cal M}_H{\cal M}^\ast_\text{cont})$.  One can choose to include the interference term in the signal: $S_i\propto\sigma(|{\cal M}_H|^2+2\,\mbox{Re}({\cal M}_H{\cal M}^\ast_\text{cont}))$.\footnote{In principle, $S_i$ can be negative, and this does affect phenomenologically relevant distributions \cite{Campbell:2011cu}.}
We assess interference effects using two measures: 
$R_1=\sigma(|{\cal M}_\text{VV}|^2)/\sigma(|{\cal M}_H|^2 + |{\cal M}_\text{cont}|^2)$ 
and $R_2=\sigma(|{\cal M}_H|^2+2\,\mbox{Re}({\cal M}_H{\cal M}^\ast_\text{cont}))/\sigma(|{\cal M}_H|^2)$.
As shown in Table \ref{tab:results}, at $\sqrt{s} = 7$ TeV interference effects can be as large as 5\%.
At $\sqrt{s} = 14$ TeV, interference effects can approach the $10\%$ level:
for $WW$ production and standard cuts (Higgs search cuts) one obtains $R_1=0.9680(8)$ ($R_1=0.940(2)$) and $R_2=0.932(2)$ ($R_2=0.926(2)$), and 
for $ZZ$ production and standard cuts (Higgs search cuts) one obtains $R_1=0.969(2)$ ($R_1=1.011(2)$) and $R_2=0.945(3)$ ($R_2=1.011(3)$).  While the additional Higgs search cuts increase the negative interference for $WW$ production, for $ZZ$ production
the $|M_{l\bar{l}l'\bar{l'}} - M_H| < \Gamma_H$ cut limits the interference effect
to about $1\%$.  The latter traces back to the fact that the interference changes
sign at $M_{l\bar{l}l'\bar{l'}} = M_H$, as seen in Fig.\ \ref{fig:ZZ} (left).

\begin{figure}
\begin{center}
\includegraphics[height=5.cm, clip=true]{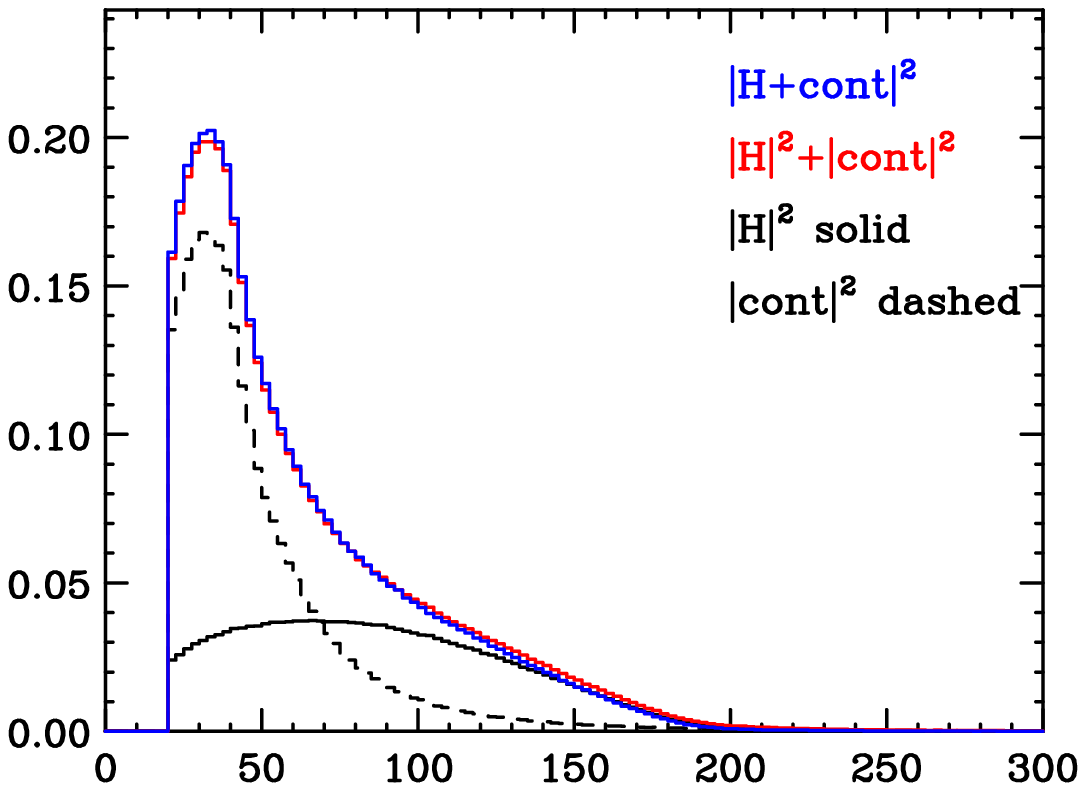}\hfil
\includegraphics[height=5.cm, clip=true]{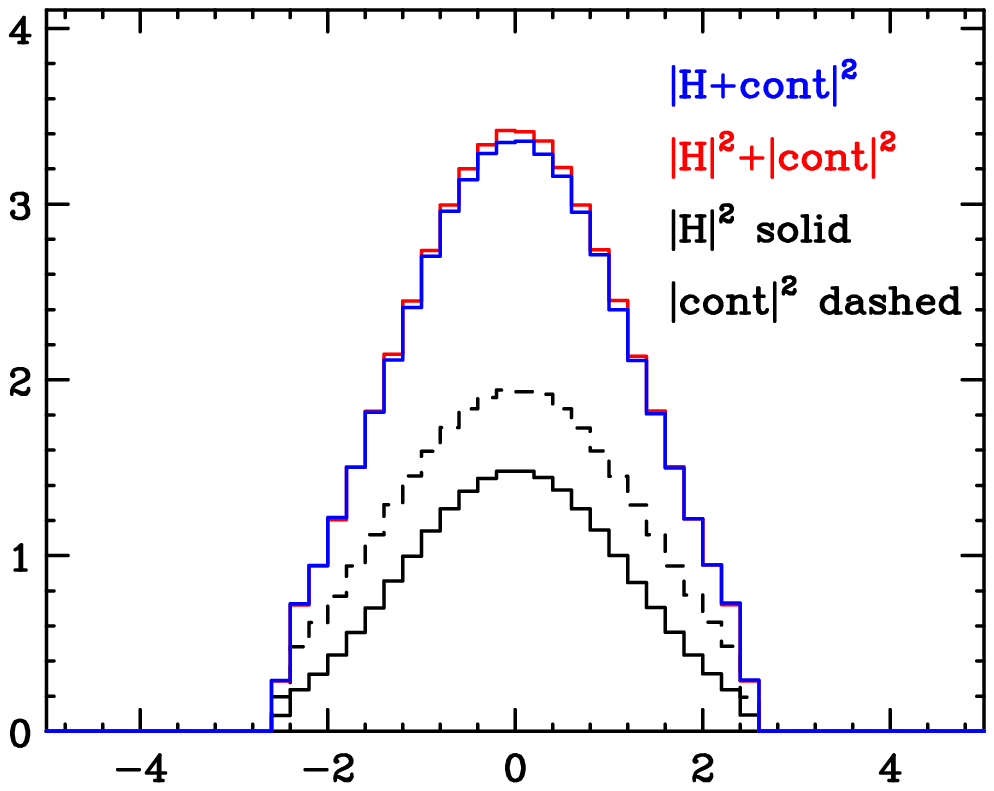}
\end{center}
\caption{Differential cross section distributions for $gg\ (\to H)\to W^-W^+\to l\bar{\nu}_l\bar{l'}\nu_{l'}$ in $pp$ collisions at $\sqrt{s} = 7$ TeV for $M_H=400$ GeV and a single lepton flavour combination calculated at LO.  The $p_{Tl}$ [GeV] (left) and $\eta_l$ (right) distributions are shown. Standard cuts are applied (see main text). fb is used as cross section unit.}
\label{fig:WW1}
\end{figure}

\begin{figure}
\begin{center}
\includegraphics[height=5.cm, clip=true]{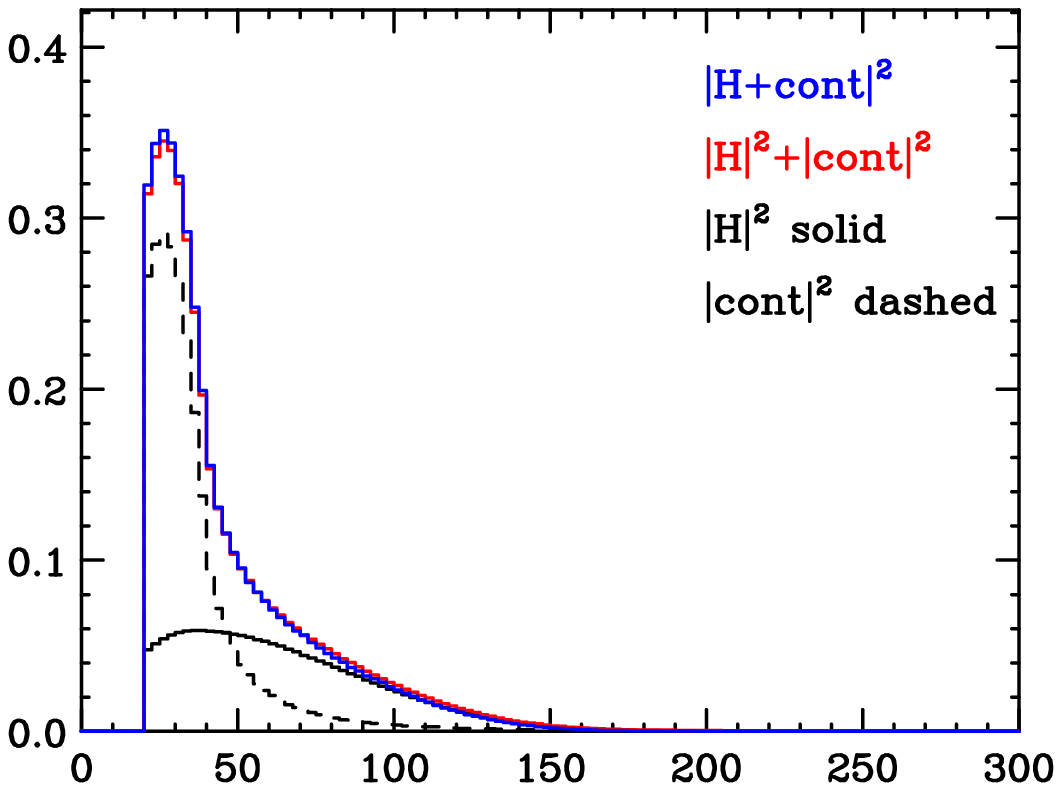}\hfil
\includegraphics[height=5.cm, clip=true]{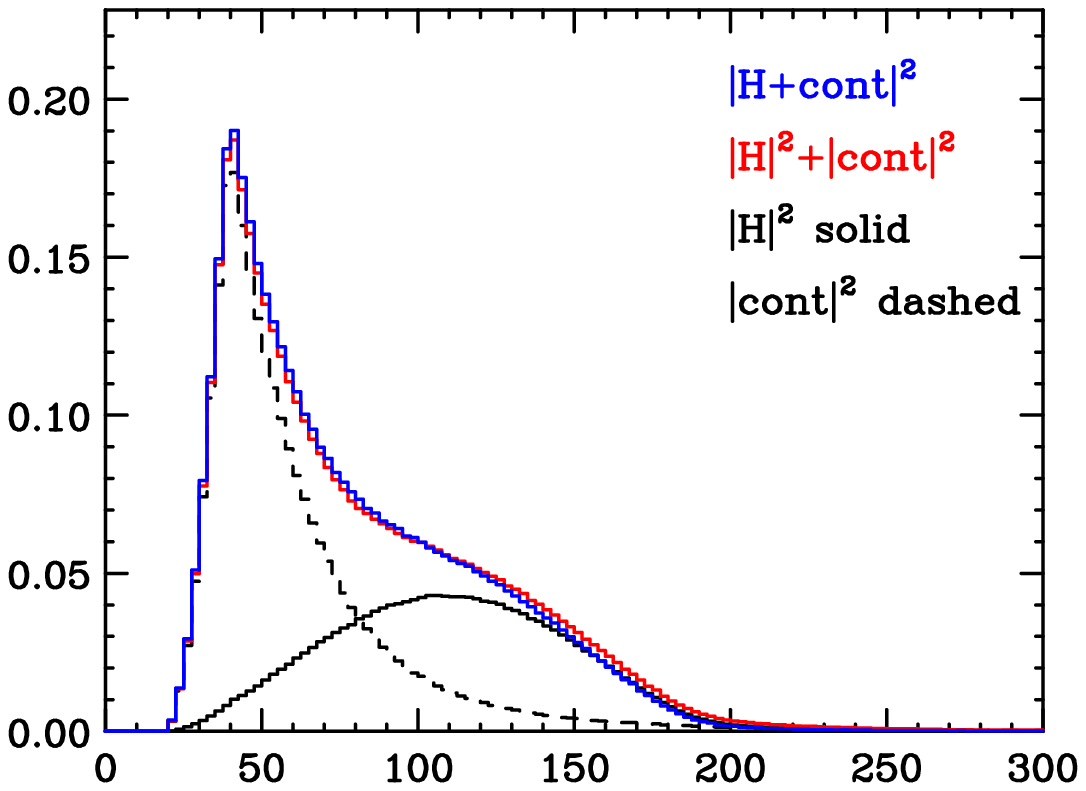}
\end{center}
\caption{$p_{Tl,\text{min}}$ [GeV] (left) and $p_{Tl,\text{max}}$ [GeV] (right) distributions.  Other details as in Fig.\ \protect\ref{fig:WW1}.}
\label{fig:WW2}
\end{figure}

\begin{figure}
\begin{center}
\includegraphics[height=5.cm, clip=true]{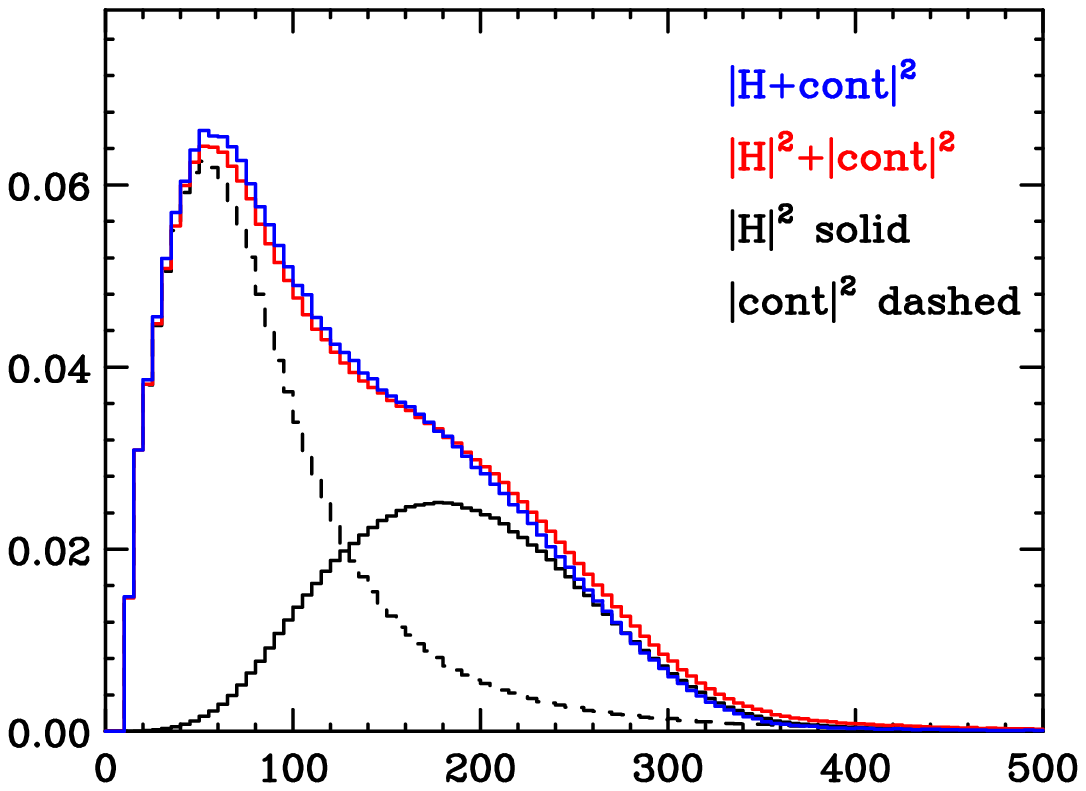}\hfil
\includegraphics[height=5.cm, clip=true]{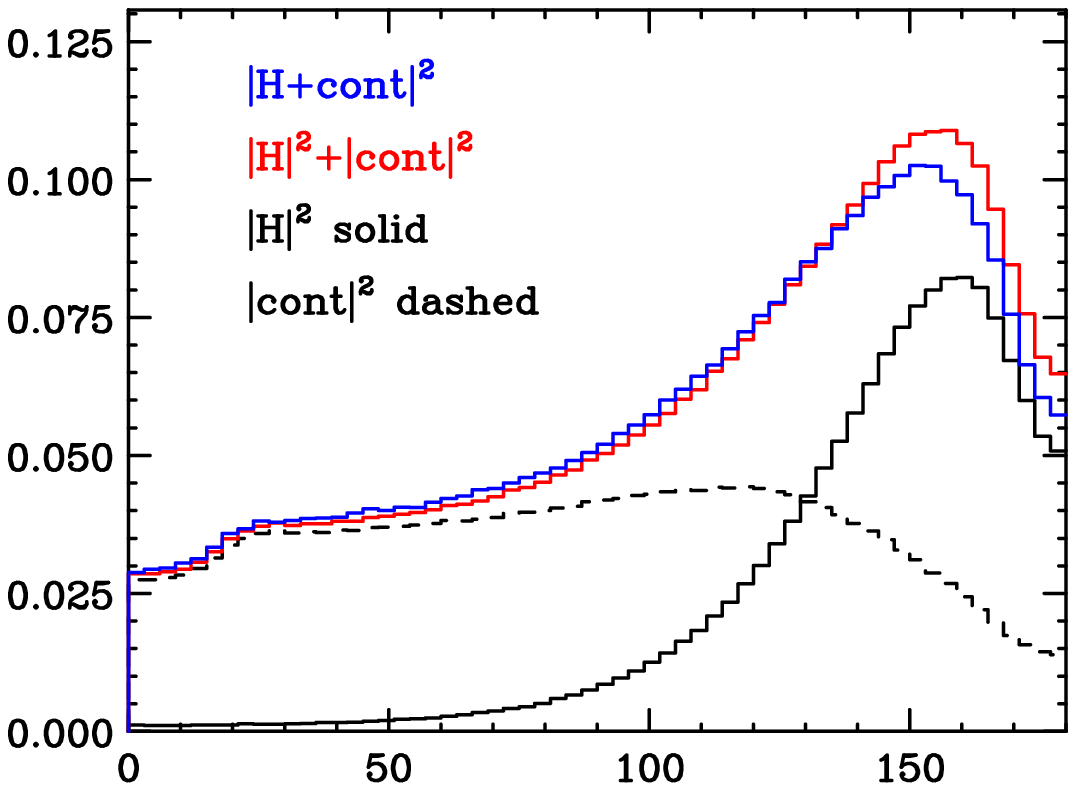}
\end{center}
\caption{$M_{l\bar{l'}}$ [GeV] (left) and $\Delta\phi_{l\bar{l'}}$ [$^\circ$] (right) distributions.  Other details as in Fig.\ \protect\ref{fig:WW1}.}
\label{fig:WW3}
\end{figure}

Differential distributions for phenomenologically relevant observables
are shown in Figs.\ \ref{fig:WW1}, \ref{fig:WW2} and \ref{fig:WW3} 
for $WW$ production
and in Fig.\ \ref{fig:ZZ} for $ZZ$ production.  The distributions demonstrate
that a compensation of positive and negative interference occurs for
cross sections that are integrated over most of the phase space.
For this reason, the more exclusive selection cuts typically used in
Higgs and new physics searches can increase the size of the interference
effects.

% =========================================================================

\section{Conclusions}

$gg\ (\to H) \to VV$ interference effects are not suppressed and can range from
1\% to about 10\%. They can be enhanced by Higgs search selection cuts and should be taken into account in the LHC data analysis.

% =========================================================================

\acknowledgments
First, I would like to thank the organisers for the invitation to speak
at the RADCOR 2011 conference. 
Helpful discussions with A.\ Djouadi and M.\ Kr\"{a}mer 
in the initial stage of this project are gratefully acknowledged.
I would also like to thank G.\ Heinrich for important discussions and  
numerical comparisons of the employed $gg\to W^-W^+\to l\bar{\nu}_l\bar{l'}\nu_{l'}$ and $gg \to Z(\gamma^\ast)Z(\gamma^\ast) \to l\bar{l}l'\bar{l'}$ amplitudes with \texttt{golem-2.0} \cite{Cullen:2011ac} generated implementations.
Helpful discussions with N.\ De Filippis, R.\ K.\ Ellis, C.\ Mariotti, P.\ Nason, S.\ Pozzorini, I.\ Puljak, T.\ Riemann, R.\ Tanaka and D.\ Zeppenfeld are also gratefully acknowledged.
This work was carried out as part of the research programme of
the Royal Holloway and Sussex Particle Physics Theory Consortium
and the NExT Institute.
Financial support under the SEPnet Initiative from the Higher Education
Funding Council for England and the Science and Technology Facilities
Council (STFC) is gratefully acknowledged.  This work was supported by 
STFC grant ST/J000485/1.

\end{document}